\documentclass[%
 reprint,
 amsmath,amssymb,
 aps, prd, longbibliography,
]{revtex4-1}

\usepackage[utf8]{inputenc}

\usepackage{graphicx}
\usepackage{dcolumn}
\usepackage{bm}

\usepackage{hyperref}
\usepackage{algorithm}
\usepackage{algpseudocode}
\usepackage{xcolor}
\usepackage{upgreek}
\usepackage{siunitx}
\usepackage{amssymb}

\usepackage{hyperref}

\hypersetup{
    colorlinks=true,
    linkcolor=black,
    citecolor=black,
    filecolor=magenta,      
    urlcolor=blue,
    pdfpagemode=FullScreen,
    }

\newcommand{\refeq}[1]{Eq.~(\ref{eq:#1})}          
          
\newcommand{\reffig}[1]{Figure~\ref{fig:#1}}

\newcommand{\rmd}{{\rm d}}

\begin{document}

\preprint{APS/123-QED}
\title{Double White Dwarf Tides From Multi-messenger Measurements}

\author{Nathaniel Leslie}
\email{nathaniel\_leslie@berkeley.edu}
\author{Liang Dai}
\email{liangdai@berkeley.edu}
\affiliation{Department of Physics\\
University of California, Berkeley, CA 94720, USA}
\author{Kevin Burdge}
\email{kburdge@mit.edu}
\affiliation{Department of Physics\\ 
Massachusetts Institute of Technology, Cambridge, MA 02139, USA}

\date{\today}

\begin{abstract}

Short-period Galactic double white dwarf (DWD) systems will be observable both in visible light through photometric monitoring and in \si{mHz}-range gravitational waves (GWs) with forthcoming space-based laser interferometry such as LISA. When only photometric variability is used to measure DWD intrinsic properties, there is a degeneracy between the chirp mass and binary tidal interaction, as orbital frequency time derivative is set by both GW radiation and tides. Without expensive radial velocity data from spectroscopic monitoring, this degeneracy may be lifted in principle by directly measuring the second time derivative of the orbital frequency through photometric monitoring over an ultra-long time baseline. Alternatively, the degeneracy can be removed by exploiting information in both photometric variability and the coherent GW waveform. Investigating both approaches, we find that direct measurement of the second time derivative is likely infeasible for most DWDs, while the multi-messenger method will disentangle measurements of the chirp mass and the binary moments of inertia, for a large sample of tidally locked systems. The latter information will enable empirical tests of WD structure models with finite temperature effects.

\end{abstract}

\maketitle

\section{Introduction}
\label{sec:intro}

Short-period double white dwarf (DWD) systems in the Milky Way are one of the loudest predicted populations and the only observationally-guaranteed population of gravitational wave (GW) sources at low frequencies $10^{-4}$--$10^{-2}\,$Hz detectable by space-based interferometric observatories such as a forthcoming leading mission LISA~\cite{lisa} (see also the TianQin mission \cite{tianqin}).
Along with GW signals from super-massive black hole binaries (SMBHBs) and extreme mass ratio inspirals (EMRIs), GW signals from DWDs are expected to be persistent and overlapping in both time and frequency, unlike the short events measured by the LIGO-Virgo-KAGRA (LVK) collaboration. The loudest of these signals are individually detectable in GWs and can be fit and removed from the data. Those include the DWDs \cite{littenberg2024lisaverification, gaialisa2018, gaialisa2024, Burdge_2020} that have already been observed electromagnetically, which will enable complementary constraints on their binary orbital motions. A much larger number of these signals will be unresolvable and they will form a stochastic GW background in the frequency range $10^{-4}$--$10^{-3}\,$Hz \cite{Cornish_2017, LISAconfusionfit}.
This confusion noise, in the particular case of a nominal four-year LISA mission and after subtraction of individually detectable signals, is predicted to be the dominant contribution to the strain power spectral density in the frequency range $0.5-2 \, \si{\mHz}$ \cite{LISAconfusionfit, multimessengerDWDprospects, lisapsd}.

It has been estimated that in the Milky Way tens of millions of DWD systems exist across the LISA frequency band \cite{Korol_DWD_population_model}, but only $\mathcal{O}(10^2)$ LISA-band DWDs have been detected in the optical domain as short-period eclipsing binaries \cite{gaialisa2024, WD_CE, 7minDWD, 8minDWD2020, rebassamansergas2024j05265934}.

So far, optical detections are made through major spectroscopic and photometric monitoring programs such as the Extremely Low Mass (ELM) survey \cite{ELM} and the Zwicky Transient Facility (ZTF) \cite{ZTF}. The Gaia mission has provided astrometry, allowing us to compute distances to sources, and will likely enable more detections when photometry is released \cite{gaialisa2018, gaialisa2024}. Over the next decade, substantially more short-period DWDs are expected to be detected first electromagnetically at the Vera C. Rubin Observatory (also known as LSST), which we consider in this work, and then with gravitational waves LISA. Measurements of both magnitude and GW strain as a function of time from the same DWD source independently allow us to extract information about the orbital phase evolution, with the GW phase being always twice the orbital phase; these two different observational avenues also yield unique and complementary information about other intrinsic and extrinsic properties of the DWD system. The combination of photometric and GW signals will enhance detections of individual DWDs and lead to more precise measurements of their stellar and orbital properties \cite{Shah_2012, Shah_2013}, and hence will help shed light on the formation and evolution of DWDs on tight orbits.



DWD inspiral in the LISA band is likely a later stage in the dynamic evolution of DWD systems. The progenitor stars are too big to fit into the current tight DWD orbits observed, while orbital decay via GW radiation alone is insufficient to allow initially widely separated DWDs to merge within a Hubble time \cite{CEreview, Peters_1964}. A promising explanation for the formation of these systems is a common-envelope (CE) phase, in which the compact cores of both progenitor stars spiral within a shared stellar envelope of hydrogen following unstable binary mass transfer \cite{Paczynski_1976}. Such CE dynamics leads to rapid ejection of the shared hydrogen envelope and the compact cores stall on a very tight orbit. The fraction of orbital energy dissipated in this process has been constrained by DWD population studies \cite{WD_CE}. Alternatively, the WD binary could have been part of a hierarchical triple system, and via the Kozai-Lidov (KL) effect \cite{Kozai, Lidov, Naoz_2016} it could have been driven onto a highly eccentric orbit with a large semi-major axis but a tiny periastron. With energy loss from tides and GW radiation, the binary orbit can circularize and settle down to a short period. Both formation channels are expected to produce systems of circular orbits in the LISA band. To our knowledge, eccentric systems with short periods, although expected from dynamic formation channel in dense stellar systems~\cite{Willems_2007}, are yet to be found.

Short-period DWDs are further tightening their binary orbits through GW radiation. When these systems reach a critical orbital period of roughly $P_c \approx 45$--$130\,$min (depending on the exact WD masses and ages) \cite{fullerlai2012, 7minDWD}, 
tidal torquing turns on to drive the WDs toward tidal locking, for which the WD spin periods are synchronized with the binary orbital period. Since WD spins open up an additional angular momentum reservoir, such tidal interaction modifies the rate of orbital tightening, and hence the orbital phase evolution measurable with both photometry and GWs. When the DWD orbit becomes sufficiently tight, one WD will overflow its Roche lobe and begin to transfer mass to the other, which further impacts the orbital evolution. This mass transfer can eventually cause the accreting object to trigger a Type Ia supernova (SN), which are used as a standard candle for measuring the expansion rate of the Universe \cite{1aSN_Perlmutter1999, 1aSN_2022}. The mechanisms through which this can happen are still being studied. Recently, the dynamically-driven double-degenerate double-detonation (D$^6$) scenario has been used to explain the existence of high speed white dwarfs in Gaia data by proposing they are the companions of the exploding star in the scenario \cite{Shen_2018, Bauer_2021}.

It is easier to isolate the effects of the tides on the orbital evolution prior to the onset of Roche-lobe overflow, so in this work we shall consider detached DWDs, which should also be more common than mass-transfer systems due to longer orbital evolution timescales. By analyzing tidal effects on the orbital decay rate, we can constrain the combined moment of inertia of the two WDs, which will allow for novel constraints on WD structure and internal physics \cite{7minDWD}. In addition, tidal effects before Roche-lobe overflow set crucial initial conditions for WD mergers \cite{Fuller_2014}.

In this paper, we turn specific attention to the degeneracy that rises between the dependence of orbital phase evolution on the tides and that on the WD masses. Some previous works assumed a value for the tidal contribution to orbital decay, but they are unable to measure it independently of the contribution from GW radiation \cite{8minDWD2020, 7minDWD}. In Section \ref{sec:tides}, we will present a framework to quantify the effect of tides on DWD orbital evolution for detached, tidally locked DWD systems. In Section \ref{sec:degeneracies}, we will discuss parameter degeneracies that arise from including these effects and how these degeneracies may be broken using multi-messenger information. In Section \ref{sec:injection}, we describe mock multi-messenger parameter inference we perform to test the feasibility of the methods of breaking the degeneracies. In section \ref{sec:results}, we present the results of our mock parameter inference study, and in Section \ref{sec:discussion}, we discuss the implications of these results. We give concluding remarks in \ref{sec:conclusion}.



\section{Tidal Effects in Detached, Tidally-Locked DWDs}
\label{sec:tides}

The orbital frequencies of LISA-band DWDs increase over time mainly due to GW radiation. However, this orbital evolution is expected to be slow ($f_{\rm GW} = 3\times 10^{-4}$--$10^{-2}\,\si{\Hz}$ and $\dot{f}_{\rm GW} = 10^{-18}$--$10^{-15}\,\si{\Hz\per\s}$) \cite{Korol_DWD_population_model}. The timescale for order-unity change in the orbital frequency is much longer than the expected time span of photometric surveys and GW observations. Therefore, we model the orbital phase evolution using the following low-order polynomial,
\begin{equation}
    \phi_{\rm orb}(t) = \phi_0 + 2\pi\,\left( f\,t + \frac{1}{2}\,\dot{f}\,t^2 + \frac{1}{6}\,\ddot{f}\,t^3 \right),
\label{eq:orbitalphase}
\end{equation}
where $f$, $\dot{f}$ and $\ddot{f}$ are phase derivatives defined at a chosen reference time. For circular inspiral, each orbital cycle corresponds to two sinusoidal GW cycles, and the GW phase is twice the orbital phase as parameterized in \refeq{orbitalphase}. The quadrupole formula predicts that due to GW radiation alone the orbital evolution depends on the chirp mass of the binary, $\mathcal{M}=(M_1\,M_2)^{3/5}/(M_1 + M_2)^{1/5}$ (where $M_1$ and $M_2$ are the masses of the binary components) \cite{Peters_1964}:
\begin{subequations}
\begin{equation}
    2\dot{f}_{\rm GW\,only} = \frac{96}{5}\,\pi^\frac{8}{3}\,\left(\frac{G\mathcal{M}}{c^3}\right)^\frac{5}{3}\,(2\,f)^\frac{11}{3},
\label{eq:gwfdotpuregw}
\end{equation}
\begin{equation}
    2\ddot{f}_{\rm GW\,only} = \frac{33792}{25}\,\pi^\frac{16}{3}\,\left(\frac{G\mathcal{M}}{c^3}\right)^\frac{10}{3}\,(2\,f)^\frac{19}{3}.
\label{eq:gwfddotpuregw}
\end{equation}
\label{eq:gwfreqderivativespuregw}
\end{subequations}

As previously mentioned, tidal effects will correct these frequency derivatives, which is more significant at short orbital periods. For DWDs that are undergoing tidal spin-up (or spin-down) but are not yet synchronized, one can parameterize the degree of tidal locking using a phenomenological tidal locking factor $\eta = \Omega_{\rm spin}/\Omega_{\rm orbit}$ as introduced in \cite{Piro2019}. This approach uses the simplifying assumption that both stars are spinning at the same orbital frequency $\Omega_{\rm spin}$. When the period of a DWD system drops below the critical orbital period $P_c \approx 45$--$130$ minutes, $\eta$ asymptotes towards 1 (complete tidal locking) as shown in Figure 12 of \cite{fullerlai2012}. As the system reaches short periods, $\eta$ changes more slowly such that $\dot{\eta}/\eta \ll \dot{f}/f$. Because of this, we will allow a general value of $\eta$, but we will use the simplifying assumption that $\eta$ is a constant in time. 

Assuming complete tidal spin-orbit synchronization, we can derive tidal corrections to the frequency derivatives. The binding energy of the binary is the sum of gravitational potential energy $E_g$ and kinetic energy of the binary orbital motion $E_{k,{\rm orb}}$:
\begin{equation}
    E_{\rm orb} = E_g + E_{k, {\rm orb}} = -\frac{G\,M_1\,M_2}{2\,a},
\label{eq:Eorb}
\end{equation}
where $a$ is the binary semi-major axis. Applying Kepler's 3rd Law, the rate of change in $E_{\rm orb}$ is proportional to the orbital period derivative
\begin{equation}
    \dot{E}_{\rm orb} = \frac{GM_1M_2}{3\,a}\,\frac{\dot{P}}{P}.
\label{eq:Eorbdot}
\end{equation}
If the finite sizes of the WDs are neglected, this energy would be lost only to GW radiation. When the finite sizes are accounted for, orbital energy partially converts into WD rotation energy, and is partially lost to tidal heating, in addition to GW loss. When tidal-locking is achieved, the tidal heating rate is very low \cite{fullerlai2012}, so we neglect this contribution and write
\begin{equation}
    \dot{E}_{\rm orb} = \dot{E}_{k, {\rm rot}} + \dot{E}_{\rm GW}.
\label{eq:Etidedot}
\end{equation}

Assuming the stars rotate at the same rate parameterized by the constant $\eta = \Omega_{\rm spin}/\Omega_{\rm orbit}$, the rotational kinetic energy of the WDs change at a rate
\begin{equation}
    \dot{E}_{k, {\rm rot}} = -4\pi^2\,I\,\frac{\dot{P}}{P^3}\,\eta^{-2},
\label{eq:Ekrotdot}
\end{equation}
where $I=I_1+I_2$ is the sum of moments of inertia of the two WDs.

Now we can use these expressions to calculate the orbital frequency derivative in terms of the power transferred into the GWs:
\begin{equation}
    \frac{\dot{f}}{f} = - \frac{\dot{P}}{P} = \frac{3\,a}{G\,M_1\,M_2}\,\frac{\dot{E}_{\rm GW}}{1-r},
\label{eq:fdottide}
\end{equation}
where $r$ is the absolute value of the ratio of Eq.(\ref{eq:Ekrotdot}) and Eq.(\ref{eq:Eorbdot}) and is given by
\begin{equation}
    r \equiv \frac{12\pi^2\,I\,a}{G\,M_1\,M_2\,P^2\eta^2}.
\label{eq:r}
\end{equation}
The period derivative without tidal effects is recovered if $I$ or $r$ are set to zero. we can write the orbital frequency derivative as
\begin{equation}
    \frac{\dot{f}}{f} = \frac{\dot{P}}{P} = \frac{\dot{f}_{\rm GW\,only}}{f}\,\left(1 + \frac{r}{1-r}\right).
\label{eq:fdottide}
\end{equation}
To determine the second frequency derivative $\ddot f$, we need to know $\dot{r}$. Applying Kepler's 3rd law to relate $a$ and $P$, we have $r \propto P^{-4/3} \propto f^{4/3}$. Here we are assuming for simplicity that $\dot{\eta}=0$. We derive
\begin{equation}
    \frac{\dot{r}}{r} = \frac{4}{3}\,\frac{\dot{f}}{f} = \frac{4}{3}\,\frac{\dot{f}_{\rm GW\,only}}{f}\,\left( 1+\frac{r}{1-r} \right).
\label{eq:rdot}
\end{equation}
We simplifying our final expressions by replacing $r$ with a different dimensionless ratio $r_{\rm tide}$, which is defined as
\begin{equation}
    r_{\rm tide} \equiv \frac{\dot{P}_{\rm tide}}{\dot{P}_{\rm GW}} = \frac{r}{1-r}.
\label{eq:rtide}
\end{equation}
The first frequency derivative can be written as
\begin{equation}
    \dot{f} = \dot{f}_{\rm GW\,only}\,\left( 1+r_{\rm tide} \right)
\label{eq:gwfdotrtide}
\end{equation}
The second frequency derivative is then derived by taking the time derivative of Eq.(\ref{eq:rdot}), using $\dot{f}_{\rm GW\,only} \propto f^{11/3}$, and applying Eq.(\ref{eq:rdot})
\begin{equation}
    \ddot{f} = \frac{(\dot{f}_{\rm GW\,only})^2}{3 \,f}\,\left( 1 + r_{\rm tide} \right)^2\,\left( 11 + 4\,r_{\rm tide} \right).
\label{eq:gwfddot}
\end{equation}

\section{Resolving Degeneracy Between Chirp Mass and Tides}
\label{sec:degeneracies}

In previous analyses of photometrically monitored DWD systems, mass measurements are carried out by constraining the time derivative of the orbital frequency \cite{8minDWD2020, 7minDWD}. With only the first frequency derivative $\dot f$ measured, there is a degeneracy between $\mathcal{M}$ and $r_{\rm tide}$ for tidally locked DWDs. If the second frequency derivative $\ddot{f}$ is measurable, this degeneracy can be lifted. Unfortunately, for typical short-period detached DWDs $\ddot{f}$ is extremely small $\ddot{f}_{GW} \lesssim 10^{-28}\,{\rm s}^{-3}$. \refeq{gwfdotpuregw} and \refeq{gwfddot} imply a steep scaling $\ddot{f} \propto f^{19/3}$, so the best hope for measuring $\ddot f$ would lie in those detached DWDs with the shortest periods ($P \lesssim 10 \,$min).
This possibility of breaking this degeneracy by measuring $\ddot{f}$, as well as the biases that arise from neglecting tidal effects are discussed for LISA data alone in \cite{fiacco2024uncoveringstealthbiaslisa}.
As we will show, measuring $\ddot{f}$ will require very long time baseline and high precision of phase measurement, which suggests that the best approach is extremely dedicated photometric monitoring.

Additional information could be exploited to break the degeneracy between $\mathcal{M}$ and $r_{\rm tide}$. Radial velocity measurements from spectroscopy can be used to independently constrain orbital velocities and thus disentangle parameter degeneracy between orbital period, orbital separation, and masses. Getting radial velocity measurements from both stars requires the stars to be extraordinarily close in temperature so that both lines are detectable. This requires extremely tight fine tuning of star temperatures, and is usually impossible. Single radial velocity measurements using the 10m Keck telescope were used to analyze bright ZTF sources in \cite{Burdge_2020}, but the information was not sufficient to break this degeneracy. This degeneracy could be broken with higher precision single radial velocity measurements, but this would require a substantial amount of dedicated time on an extremely large telescope, which is often cost prohibitive.

Alternatively, a coherent GW signal detected will provide this information. At a given orbital frequency $f$, the observed strain amplitude depends on the chirp mass $\mathcal{M}$, the luminosity distance to the source $d_L$, and the orbital inclination $\iota$, but is does not depend on $r_{\rm tide}$.
For eclipsing DWDs, $\iota$ will be well determined by analysis of the electromagnetic (EM) lightcurve. The luminosity distance $d_L$ can be extracted from the photometric magnitudes of the binary if temperature and radii are independently constrained by full analysis of the lightcurve.
For an eclipsing DWD system, surface temperatures can be measured with multi-band photometry and the radii of the WDs can be measured from the eclipse signals.

In this work, we would like to compare two methods to break the degeneracy between $\mathcal{M}$ and $r_{\rm tide}$: one approach by directly measuring $\ddot{f}$ and the other multi-messenger approach by combining photometric and GW signals. Many eclipsing DWDs are or will be initially discovered in large-sky-coverage photometric cadence surveys, and for this reason, we consider the Vera C. Rubin Observatory (also known as LSST), a six-band flagship program which will survey roughly half of the sky over a decade \cite{lsstsciencebook}. By the time LISA is operational, this 10-year survey will have been carried out, and many eclipsing DWDs individually detectable with LISA will have their optical counterparts identified in LSST. However, LSST will have a 30-second exposure time for each image, which can partially smear out the eclipsing signal, whose duration can be as short as a few minutes. To explore the best chance at detecting $\ddot{f}$ with a good precision, we also consider HiPERCAM, an existing short-cadence camera that has been deployed on the 10.4-m Gran Telescopio Canarias (GTC) \cite{HiPERCAM}. In order to balance readout noise and shot noise, we assume 3 second visits as was done for HiPERCAM in \cite{Burdge_2020}.

\section{Mock Parameter Inference}
\label{sec:injection}

To investigate how effective the multimessenger approach and photometric $\ddot{f}$ measurements with HiPERCAM are for breaking the $\mathcal{M}$--$r_{\rm tide}$ degeneracy, we generate mock photometric and GW data with injected DWD signals, and perform mock Bayesian parameter estimation to derive posterior distributions for the source parameters. In this section, we will explain how we generate mock data for LISA, LSST, and HiPERCAM, respectively.

\subsection{LISA}
\label{sec:lisa}

LISA will consist of three spacecrafts in triangular formation that orbit the sun in a cartwheeling heliocentric orbit that trails the Earth by about 20 degrees \cite{lisa}. Each spacecraft is equipped with two lasers that can beam signals to each of the other two spacecrafts. Each spacecraft is also equipped with detection systems that can measure the phases of the incoming laser signals from the other two spacecrafts. There are a total of six time series that result from measuring the residual phase of each laser signal between emission and detection. 

With onboard instrumentation accounting for most of the noise sources, the dominant noise source is laser frequency noise, which is different for each of the six lasers. This laser frequency noise can be canceled by using linear combinations of these time series, each delayed by different amounts. This process, called Time Delay Interferometry (TDI) \cite{lisatdi}, gives us 3 time series that minimize noise, called $X(t)$, $Y(t)$, and $Z(t)$, which cyclically transform into each other under cyclic permutation of the satellites.

These three time series have correlated noises. The noises can be made uncorrelated by diagonalizing the covariance matrix and finding the eigenvectors that describe linear combinations of $X(t)$, $Y(t)$ and $Z(t)$. These linear combinations, $E(t)$, $A(t)$ and $T(t)$, are given by

\begin{subequations}
\begin{equation}
    E = (X - 2\,Y + Z)/\sqrt{6},
\label{eq:E}
\end{equation}
\begin{equation}
    A = (Z - X)/\sqrt{2},
\label{eq:A}
\end{equation}
\begin{equation}
    T = (X + Y + Z)/\sqrt{3}.
\label{eq:T}
\end{equation}
\label{eq:EAT}
\end{subequations}

Due to full symmetry in this definition, no astrophysical source will produce signals in the $T$ channel. The time-domain expressions we use for $X$, $Y$, and $Z$ are detailed in Section 8.3 of \cite{ldcmanual}. We use Solar System Barycenter (SSB) conventions from Section 6.1, spacecraft orbits from Section 8.1, and a cubic polynomial for the phase that is twice of Equation \ref{eq:orbitalphase}. We assume that the three channels have stationary Gaussian noise described by power spectral densities (PSDs) $S_E(f)=S_A(f)$ and $S_T(f)$ as defined in Section 8.3, with further optical metrology and acceleration noise estimates defined in \cite{lisapsd}.

The assumption of stationary Gaussian random noise is an ideal one. The assumption of stationarity may not be accurate for several reasons; for example, the rotation of the LISA constellation plane relative to the Galactic plane, from which the bulk of the Galactic DWD confusion noise originates, will cause temporal variations in the total noise power spectra \cite{timevaryinglisanoise}. We also assume that the PSDs are known to infinite precision, while a more accurate likelihood model would marginalize over PSD uncertainties \cite{psduncertainty}. However, we lack further detailed information to account for these complications as neither real LISA data nor extensive survey data on the bulk of the Galactic DWD population is available. Additionally, as LISA will not launch for at least another decade, our PSD estimates are based on design specifications to be realized, so for this work, we will settle for stationary Gaussian noises with PSDs taken from the LISA design specifications as provided in \cite{lisapsd}.

With the assumption of stationary noise, we would need to evaluate LISA waveforms in the frequency domain for DWDs. To allow efficient computation, we use a heterodyned frequency domain waveform similar to that described in Appendix A of \cite{fdlisaheterodyne}. Essentially, we divide the complex-valued frequency-domain signal by a monochromatic complex exponential factor with the frequency of the GW at the reference time
which yields a term that varies slowly and smoothly with the frequency and contains information about frequency derivatives induced by both the intrinsic source chirping and the Doppler effect caused by the motion of the LISA constellation. This slowly varying piece can be well approximated through a computationally-efficient FFT performed on a coarse frequency grid.

To generate mock data, we assume that LISA begins to record data on January 1st, 2037 at midnight and runs for exactly 4 years at a sampling rate of 0.1 $\si{Hz}$. 

\subsection{Lightcurve Model}
\label{sec:lcmodel}

For this work, we employ a simple photometric lightcurve model for eclipsing DWDs, which accounts for all of the important effects.
We will apply this model to both LSST and HiPERCAM, except that the photometric noise level and cadence are telescope-specific, and are specified in Sections \ref{sec:lsst} and \ref{sec:hipercam}. We model four effects on the lightcurves: the spectral energy distribution for single WDs (excluding the radiation effect from the companion star), eclipses, ellipsoidal variation due to tidal deformation of the stars, and irradiation of each star by the other. For simplicity, these effects are treated as additive in the photometry.

First, we need to calculate the flux of a WD star in any given photometric filter without corrections due to the companion star. For this, we compute the fractional frequency-averaged spectral flux density $f_b$ for each band under the assumption that each WD has a blackbody spectrum \cite{freq-averaged_spectral_flux_density}. We assume an ideal filter throughput curve as defined by perfect transmission between a pair of cutoff wavelengths in Table 2.1 of \cite{lsstsciencebook}. Details such as atmospheric absorption, imperfect reflectivity/transmission for various optical components, and sensor efficiency create a more complicated transmission function that is projected to range roughly from 30-70\%, but we ignore this for simplicity.
We use the same cutoffs for both LSST and HiPERCAM. We define $f_b(T)$ to be the fractional frequency-averaged spectral flux density in the given photometric filter $b$ for a WD of a surface temperature $T$, which we can calculate as 
\begin{equation}
    f_b(T) = \frac{15\,h^3\,c^3}{\pi^4\,(k_B\,T)^4}\frac{\int_{\lambda_{{\rm min}, b}}^{\lambda_{{\rm max}, b}}\frac{\lambda^{-4}}{\exp(hc/\lambda\,k_BT)-1}\,\rmd\lambda}{\int_{\lambda_{{\rm min}, b}}^{\lambda_{{\rm max}, b}}\,\rmd\lambda/\lambda}.
\label{eq:bandfrac}
\end{equation}
The total flux of star $i$ in band $b$ is given by this multiplied by the total flux:
\begin{equation}
    \Phi_{i,b} = \frac{\sigma\,T_i^4\,R_i^2}{d_L^2}f_b(T),
\label{eq:starflux}
\end{equation}
where $T_i$ and $R_i$ are the surface temperature and radius of star $i$, $d_L$ is the luminosity distance to the star and $\sigma$ is the Stefan-Boltzmann constant.

Eclipses create two unequal dips corresponding to two transits per orbit: the primary eclipsing the secondary and vice versa. To model these dips, we borrow a simple analytic result derived for planet transits in Section 3B of \cite{robnik_lightcurve} to model flux reduction during WD transits. To compute the linear and quadratic limb-darkening coefficients, we interpolate the data in ``tableab'' associated with \cite{limb_darkening}, using the DB atmosphere for each of the $u$, $g$, $r$, $i$, $z$ and $y$ SDSS bands, which correspond reasonably well to the target LSST bands. Finally, we assume zero eccentricity and compute the eclipse durations (measured in units of the orbital phase) for given stellar radii and binary orbital inclination. These first two effects alone produce the following band-dependent lightcurve as a function of phase:
\begin{align}
    L_{\rm ecl}(\phi) &= \frac{\sigma\,T_1^4\,R_1^2}{d_L^2}\,f_b(T_1)\,\left(1 - S_1(\phi)\right)\nonumber\\ 
    &+ \frac{\sigma\,T_2^4\,R_2^2}{d_L^2}\,f_b(T_2)\,\left( 1 - S_2(\phi) \right),
\label{eq:Lecl}
\end{align}
where $S_i$ is the fractional loss in flux for the star $i$ eclipsed by the other star. At phase $\phi=0$, star 1 eclipses star 2 and $S_2$ reaches its peak value. At phase $\phi=\pi$, star 2 eclipses star 1 and $S_1$ reaches its peak value. The functional forms of $S_1$ and $S_2$ depend on both masses $M_1$ and $M_2$, radii $R_1$ and $R_2$, the orbital period $P_{\rm orb}$, and orbital inclination $\iota$. This dependence is not explicitly spelled out for simplicity of notation.

Next, we consider tidal deformation of the stellar shape. This causes the otherwise circular photosphere to apppear to the observer as an ellipse whose shape varies with the orbital phase. Such ellipsoidal variation induced by tides is often modeled directly as a sinusoid with an amplitude that depends on orbital inclination, stellar radii, binary mass ratio, as well as limb darkening and gravity darkening coefficients \cite{ellvar1993, ellvar2012, multimessengerDWDprospects}. This approximation has been shown to be inconsistent with radial velocity and measured Doppler beaming effects in calculating the binary mass ratio $q=M_2/M_1$ in the analysis of KOI-74 from Kepler data \cite{q_inconsistency}. This expression has differing values for each star and there is also band dependence on the limb darkening and gravity darkening effects. To absorb the above modeling complication, we opt to introduce a phenomenological dummy mass ratio parameter $q_{\rm dummy}$ in place of the physical mass ratio $q$ but still use this common expression. Adding this effect, the lightcurve becomes
\begin{align}
    &L_{{\rm ecl}, {\rm ell}}(\phi) = \nonumber\\ &\frac{\sigma\,T_1^4\,R_1^2}{d_L^2}\,f_b(T_1)\,(1 - S_1(\phi))\,(1 - \ell_{1,b}\,\cos(2\phi))\nonumber\\ 
    &+ \frac{\sigma\,T_2^4\,R_2^2}{d_L^2}\,f_b(T_2)\,(1 - S_2(\phi))\,(1 - \ell_{2,b}\,\cos(2\phi)),
\label{eq:Lell}
\end{align}
where $\ell_{i,b}$ is given by
\begin{equation}
    \ell_{i,b} = \frac{3\,(15+u_{i,b})(\tau_{i,b}+1)}{20\,(3-u_{i,b})}\,\left(\frac{R_i}{a}\right)^3\,q_{\rm dummy}\,\sin^2 \iota.
\label{eq:elli}
\end{equation}
In the above expression, $a$ is the binary semi-major axis, $\iota$ is the orbital inclination, and $u_{i,b}$ and $\tau_{i,b}$ are the linear limb darkening parameter and gravity darkening parameter, respectively, for star $i$ and photometric band $b$. As discussed in the previous paragraph, $q_{\rm dummy}$ is a phenomenological dummy variable used in place of the physical mass ratio $q$. We use the same linear limb darkening parameters as used in the transit expressions, and we calculate the gravity darkening coefficients using Equation 10 of \cite{limb_darkening}, using $\beta=0.25$ and the center of each photemetric band as the observed wavelength.

It is important to note that the inclination is used to compute the duration of the eclipse, which factors into the functional forms of $S_1(\phi)$ and $S_2(\phi)$. The inclination $\iota$ affects the eclipse duration, so it is not degenerate with $q_{\rm dummy}$.

\begin{figure}[!]
\centering
    \includegraphics[width=85mm]{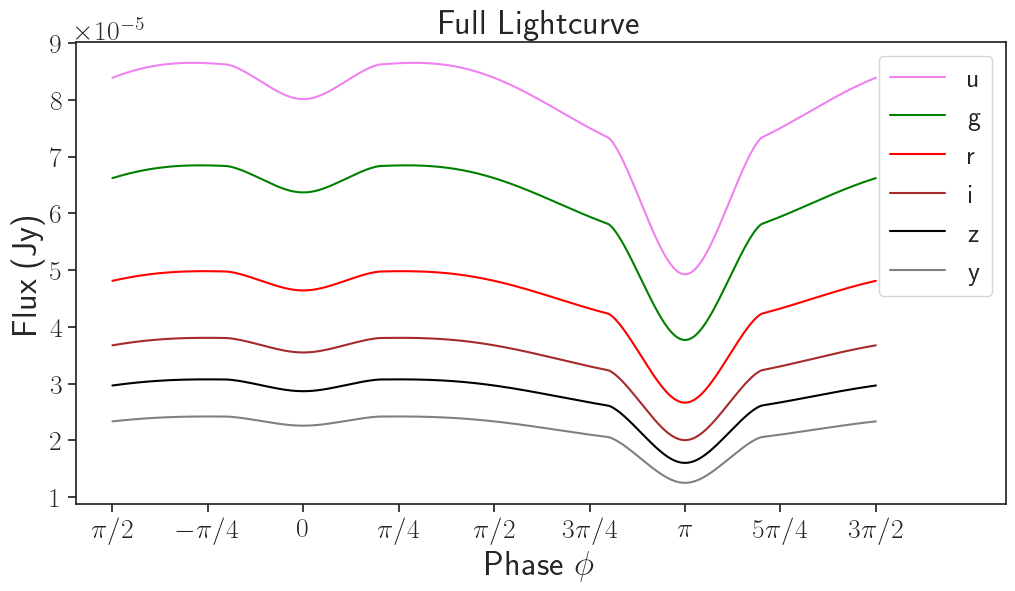} 
    \caption{Phase-folded model lightcurves from Equation \ref{eq:Lfull} using the parameters in Table \ref{tab:params} across the 6 photometric bands used by LSST. The effects of eclipsing and irradiance are clearly visible, but the effect of the ellipsoidal variation (which has 2 periods in one orbit) is difficult to discern as its amplitude is small ($\ell_{i,b} \sim 0.03-0.04$).}
    \label{fig:lightcurves}
\end{figure}

Finally, the photometric effect of WDs irradiating each other needs to be accounted for. Strictly speaking, this effect is not independent of the other effects we have considered as it creates temperature variation on the WD surface. Codes such as \texttt{ellc} have been developed to precisely calculate the full effect of this \cite{ellc}. Crudely speaking, this effect ends up looking very similar to a sinusoid with the same period as the orbit \cite{7minDWD}. In order to simply account for the degeneracy that this effect has with the others, we use this approximation with a simple phenomenological amplitude $A_{\rm irr}$ to complete our simple lightcurve model:
\begin{align}
     &L_{\rm full}(\phi) = \nonumber\\ &\left[\frac{\sigma\,T_1^4\,R_1^2}{d_L^2}\,f_b(T_1)\,(1 - S_1(\phi))\,(1 - \ell_{1,b}\,\cos(2\phi))\right.\nonumber\\ 
    &+ \left.\frac{\sigma\, T_2^4\,R_2^2}{d_L^2}\,f_b(T_2)\,(1 - S_2(\phi))\,(1 - \ell_{2,b}\,\cos(2\phi))\right]\nonumber\\
    &\times \left( 1+A_{\rm irr}\,\cos(\phi) \right).
    \label{eq:Lfull}
\end{align}
Full phase-folded lightcurves using this simple lightcurve model are shown in Figure \ref{fig:lightcurves}.

\subsection{LSST}
\label{sec:lsst}

The Vera C. Rubin Observatory (also known as LSST) hosts a $8.4$-meter (effectively $6.5$-meter) telescope which aims to survey roughly half of the sky in six broad photometric bands over a decade \cite{lsstsciencebook}. By the time LISA is launched, this 10-year survey will have been completed, and many LISA GW signals from DWDs will have optical counterparts detected with LSST. We consider a cadence for LSST that is based on the \texttt{rubin\_sim} code \cite{rubinsim} which contains a scheduler that simulates sequential decisions of which filter to use and which direction to point for the duration of LSST's operation \cite{LSST_scheduler_paper}. The scheduler also determines the 5$\sigma$ magnitude limit $m_5$ for each visit, which we use for including the photometric noise in mock LSST lightcurves. Using these values of $m_5$, the noise is modeled following Section 3.5 of \cite{lsstsciencebook}, using in-text or table values for $\sigma_{sys}$ and $\gamma$. We arbitrarily select the LSST-surveyed point (RA, DEC) = (0, $-20^{\circ}$) and we extract all of the times that this point is visited in each of the six photometric bands over a 10-year simulated survey at the expected LSST cadence. Each visit has a 30-second exposure, so our lightcurve model is time-averaged over 10 equally spaced points (each separated by 3 seconds) to account for smearing of the light variablity due to finite exposure times. If we simply compared an analysis including LSST to one including both LISA and LSST, the latter would have a longer baseline, which would not reflect true multimessenger advantages. In order to use comparable baselines for the comparison, we augment the LSST cadence with a four year chunk of the original 10 inserted while LISA flies. Whether or not LSST or other future surveys concurrent with LISA will be running, we want to make a conservative comparison as to not overstate the improvements from a multimessenger analysis. A mock example LSST cadence is shown in the upper panel of Figure \ref{fig:cadences}.

\subsection{HiPERCAM}
\label{sec:hipercam}

\begin{figure*}[!]
\begin{tabular}{c}
  \includegraphics[width=160mm]{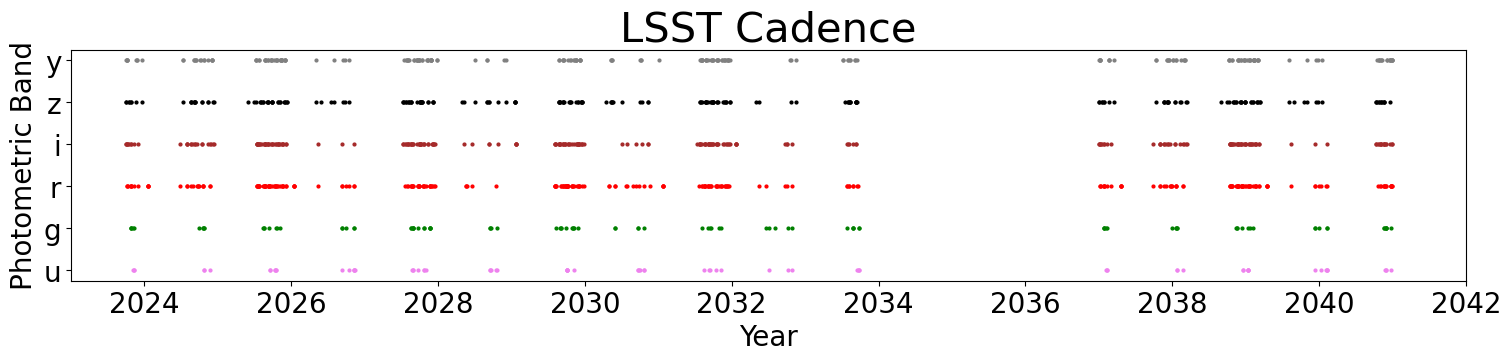} \\   \includegraphics[width=160mm]{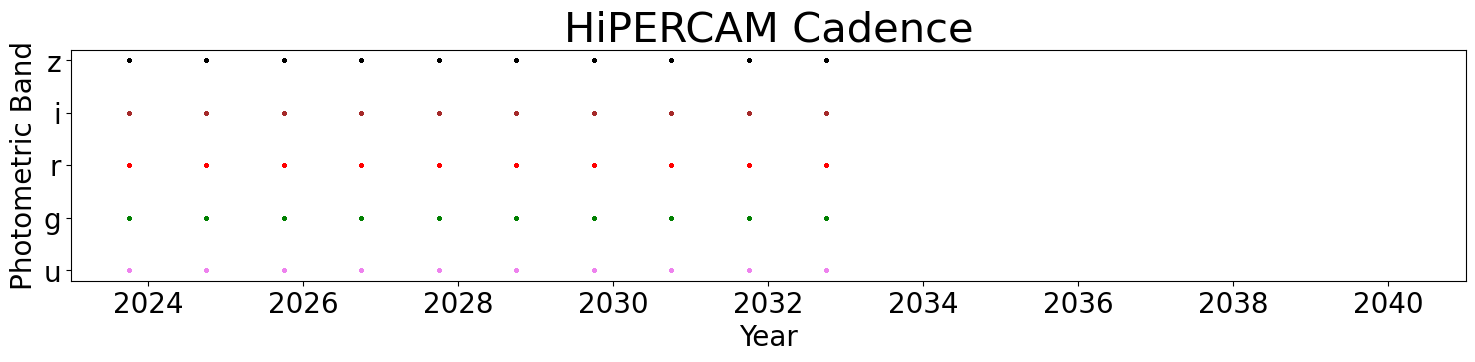}
\end{tabular}
\caption{Mock cadences for the two photometric cadence surveys we consider. For LSST (upper panel), each point represents a $30\,\si{s}$ exposure and is generated by averaging over 10 consecutive photometric measurements separated by $3\,\si{s}$ between neighboring measurements. The total exposures in seconds are 76, 100, 294, 285, 237, and 247 in the $u$, $g$, $r$, $i$, $z$ and $y$ photometric bands, respectively. For HiPERCAM (lower panel), each point consists of 1200 consecutive 3-second exposures that are simultaneously taken in all six bands.}
\label{fig:cadences}
\end{figure*}

Our mock HiPERCAM cadence begins with the first LSST measurement of the target patch of sky and consists of hour-long clusters of 3-second exposures once a year for 10 years. HiPERCAM simultaneously records data in the $u$, $g$, $r$, $i$ and $z$ photometric bands, so each $3\si{s}$ observation gathers 5 data points \cite{HiPERCAM}. Photometric noise for HiPERCAM is modeled based on real HIPERCAM data. Details of this are given in Section \ref{sec:hipercamnoise} of Appendix \ref{sec:hipercamcalibrations}. Since HiPERCAM observations are significantly shorter than the typical period of the DWDs we concern ($3\,\si{s}$ compared to $6\,\si{min}$), we choose not to include lightcurve time smearing. A mock example of the HiPERCAM cadence is shown in the lower panel of Figure \ref{fig:cadences}.


\section{Mock Parameter Inference Results}
\label{sec:results}

We investigate the efficacy of LISA and HiPERCAM in breaking the $\mathcal{M}/r_\text{tide}$ degeneracy by performing injections and generate Bayesian posteriors. We do this always with mock LSST data, and with and without both LISA and HiPERCAM mock data. Our injection parameters are based on the source ZTF J2243+5242 identified in \cite{8minDWD2020}, an eclipsing DWD system with an 8.8-minute orbital period that has already been photometrically observed. 

We use unpublished data from a 1.5 hour HiPERCAM observation of ZTF J2243+5242 provided by one of the authors, to choose the radii, inclination, and irradiation amplitude. The details of how we do this are in Section \ref{sec:hipercamdataparams} of Appendix \ref{sec:hipercamcalibrations}. We use the SED fit in \cite{8minDWD2020}, listed in their Table 3, to choose the masses, temperature, and distance. As $\ddot{f} \propto f^{19/3}$, we want to analyze a system with a very short orbital period so that it would be the most optimistic case for an attempt to directly measure $\ddot f$. We choose a reduced orbital period of 6 min; further decreasing it to 5 min would result in Roche-lobe overflow. We calculate $r_{\rm tide}$ using Equations \ref{eq:r} and \ref{eq:rtide}, evaluating each moment of inertia as $I_i = \kappa_i M_i R_i^2$, where $\kappa_1 = \kappa_2 = 0.12$ (like in \cite{8minDWD2020}) and we use $\eta=1$, as this orbital period is substantially smaller than the critical period for tidal locking of $45$--$130\,$min. In \cite{7minDWD}, the WDs in ZTF J1539+5027 with a period of 6.91 minutes were assumed to be fully tidally locked with $\eta = 1$, so we find this to be a reasonable assumption. Finally, we choose the roll angle and initial orbital phase arbitrarily, as they only depend on orientation of the source. All of our parameter choices are listed in Table \ref{tab:params}.

\begin{figure*}[!]
\centering
    \includegraphics[width=\textwidth]{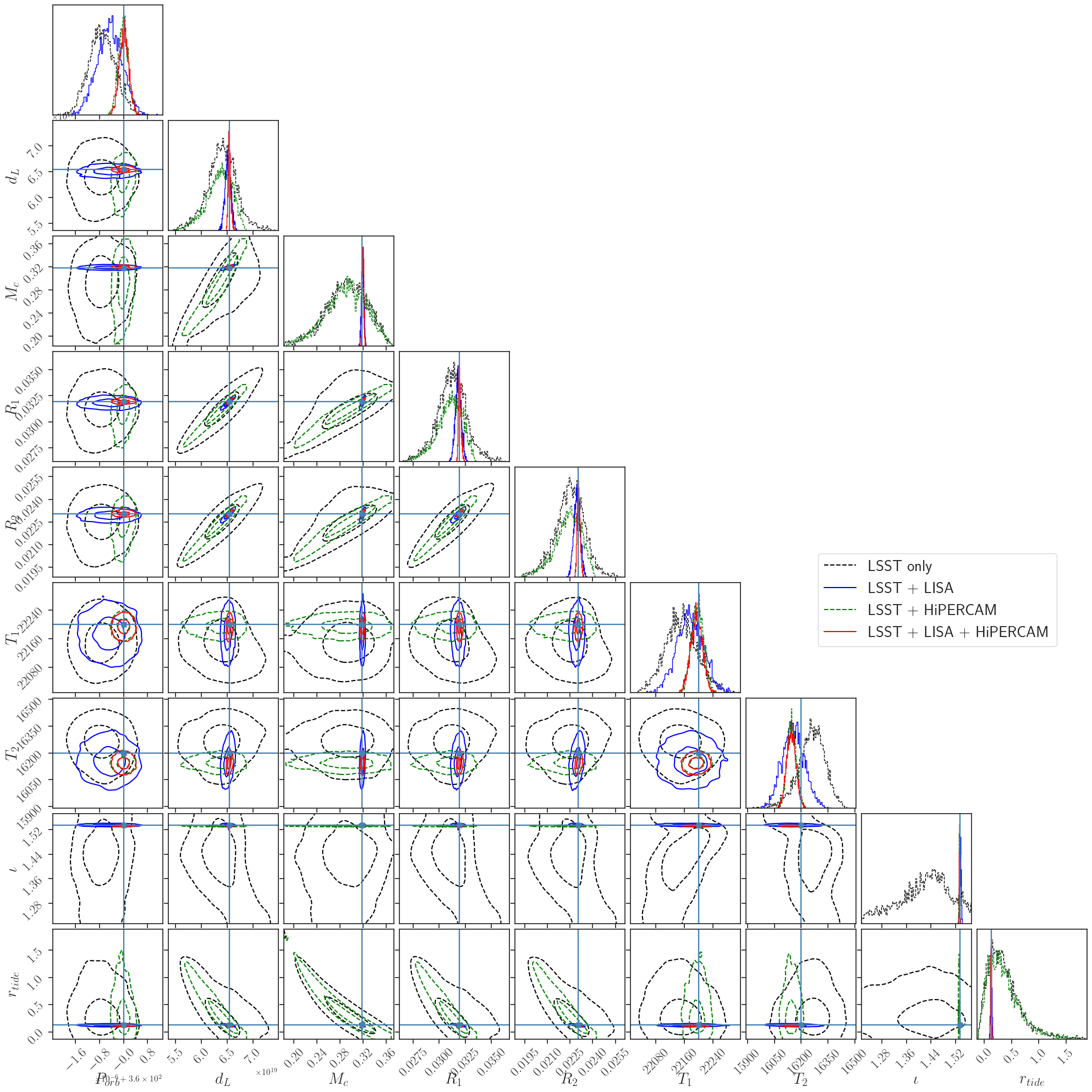} 
    \caption{Corner plot showing parameter posteriors corresponding to all four sets of detectors/telescopes. The contours shown are for 50\% and 95\% credible regions. The light blue crosses are the injected values of the parameters. The posteriors including LISA are shown with solid lines and the posteriors without LISA are shown with dashed lines. The degeneracies are significantly reduced with LISA information, suggesting that $\ddot{f}$ is not constrained properly by the EM measurement of this source.}
    \label{fig:posteriors}
\end{figure*}

While we always include LSST mock observations, we would like to compare the cases with or without LISA GW data, and with or without HiPERCAM, which results in a total of four parameter posterior distributions to be compared with each other. Bayesian parameter inference is carried out with the fast and embarrassingly parallel \texttt{pocoMC} sampler \cite{pocoMC}. We sample in mostly uniform priors, including using the $(\mathcal{M},\,q)$ basis for masses, but we apply a conservative mass-radius relation prior based on \cite{soares2017constraining}. Our prior was a Gaussian in the radius with a mean equal to the radius predicted from mass and temperature and a standard deviation equal to the difference between the mean and the the radius predicted from the same mass and zero temperature. Posterior inference for this problem was only made possible using the parameterization described in Appendix \ref{sec:reparameterization}. All four posterior distributions are plotted in Figure \ref{fig:posteriors} for a clear comparison.

\addtolength{\tabcolsep}{6pt}
\begin{table}[!]
\caption{\label{tab:params}
The injection parameters used in this work, based on our inference of HiPERCAM observations of ZTF J2243+5242 and the inference in \cite{8minDWD2020} with a orbital period shortened to 6 minutes. They list their measured parameters in Table 3. Parameters below the dividing line are not independent, and are calculated based on the above parameters, but provided for reference. We calculate $r_{\rm tide}$ using Equations \ref{eq:r} and \ref{eq:rtide}, evaluating each moment of inertia as $I_i = \kappa_i M_i R_i^2$, where $\kappa_1 = \kappa_2 = 0.12$ (like in \cite{8minDWD2020}). All priors are uniform except for masses and radii. Mass priors are uniform in chirp mass $\mathcal{M} = (M_1 M_2)^\frac{3}{5}/(M_1+M_2)^\frac{1}{5}$ and mass ratio $q = M_2/M_1$. Radius priors are Gaussian based on \cite{soares2017constraining}. The mean was the nonzero temperature prediction for the radius and the standard deviation was the difference between the zero and nonzero temperature predictions for the radius.} 
\begin{ruledtabular}
\begin{tabular}{c|c|c}
Parameter & Symbol & Injected Value \\
\hline
Orbital Period & $P_{\text{orb}}$ & $ 360 \,\si{s}$ \\
Primary Mass & $M_1$ & $0.349 \,M_\odot$ \\
Secondary Mass & $M_2$ & $0.384 \,M_\odot$ \\
Primary Radius & $R_1$ & $0.0319 \,R_\odot$ \\
Secondary Radius & $R_2$ & $0.0230 \,R_\odot$ \\
Primary Temperature & $T_1$ & $22000 \,\si{K}$ \\
Secondary Temperature & $T_2$ & $16200 \,\si{K}$ \\
Luminosity Distance & $d_L$ & $2120 \,\si{pc}$ \\
Inclination Angle & $\iota$ & $87.88^\circ$ \\
Roll Angle & $\psi$ & 0.2 \\
Initial Phase\footnote{The initial orbital phase at the first LSST observation.} & $\phi_0$ & 0.3 \\
Irradiation Amplitude & $A_{\rm irr}$ & $-0.013517$ \\
\hline
Chirp Mass & $\mathcal{M}$ & 0.3186\,$M_\odot$ \\
Mass Ratio & $q$ & 1.1002 \\
Tidal Fraction & $r_{\rm tide}$ & 0.1288
\end{tabular}
\end{ruledtabular}
\end{table}
\addtolength{\tabcolsep}{-20pt}

For this mock system, it is clear from \reffig{posteriors} that we are unable to break the degeneracy between $\mathcal{M}$ and $r_{\rm tide}$ unless GW information is available. As mentioned in Section \ref{sec:degeneracies}, this is made possible because the amplitude of the GW gives an independent measurement of $\mathcal{M}$ that does not depend on $r_{\rm tide}$.

When LISA GW detection is available, uncertainties on $\mathcal{M}$ and $r_{\rm tide}$ decrease by large factors of 24 and 37 respectively when including HiPERCAM data and decrease by factors of 17 and 25 respectively with only LSST. We see significant tightening in luminosity distance and radii as well. The uncertainty in luminosity distance $d_L$ and WD radii $R_1$ and $R_2$ all improve roughly by a factor of 9 with HiPERCAM and a factor of 5 without.

When HiPERCAM observations are available in addition to LSST observations, we see significant improvements in some of the parameters: binary period $P_{\rm orb}$, surface temperatures $T_1$ and $T_2$, and inclination $\iota$. HiPERCAM is able to achieve these improvements for 2 reasons. First, its short cadence observations which allow it to measure lightcurve details with less blurring, particularly improving the period and period derivative measurements. Second, its focused observations allow for a larger number of data points. We also note synergistic improvements with LISA as mentioned in the previous paragraph. We see even greater improvements in $\mathcal{M}$, $r_{\rm tide}$, $d_L$, $R_1$ and $R_2$ when supplementing LISA data to LSST and HiPERCAM data than we see from simply combining LISA data and LSST data.

When LISA GW detection is added to photometric observations, we are able to empirically measure $r_{\rm tide}$ without making any assumptions about moment of inertia or whether tidal locking is realized. On the other hand, knowledge of both is required in order to predict a value for $r_{\rm tide}$. If we assume that the mock DWD system is tidally locked, then we further need to know the combined moment of inertia $I = I_1 + I_2$ to determine $r_{\rm tide}$. There exist universal relations (roughly independent of composition) for high mass, low-temperature WDs that can be used to compute the moment of inertia from mass in \cite{Wolz_2020}. Our injected value of $r_{\rm tide}$ corresponds to nonzero temperature WDs as estimated in \cite{8minDWD2020} and we measure it with sufficient precision to clearly distinguish it from the prediction of these universal relations that ignore non-zero temperature effects. 

\begin{figure}[!]
    \includegraphics[width=85mm]{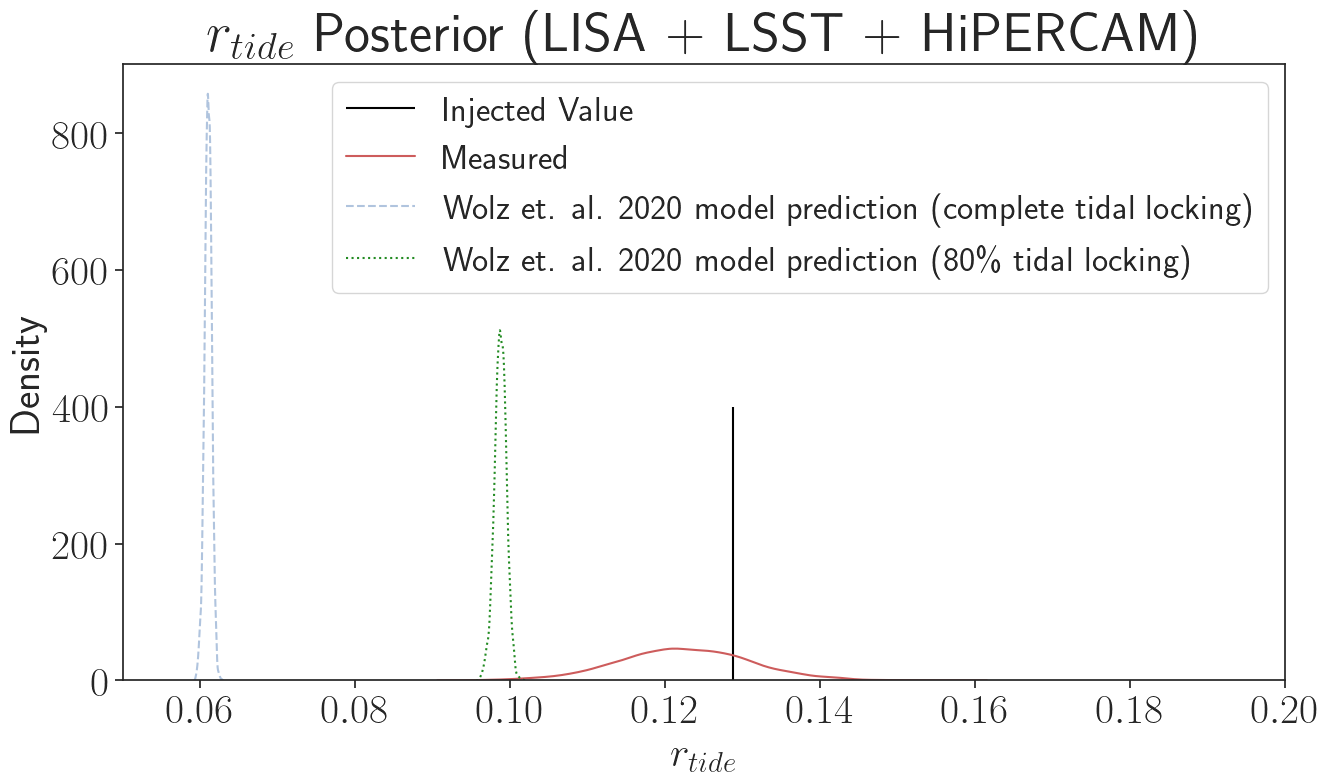} \\ \includegraphics[width=85mm]{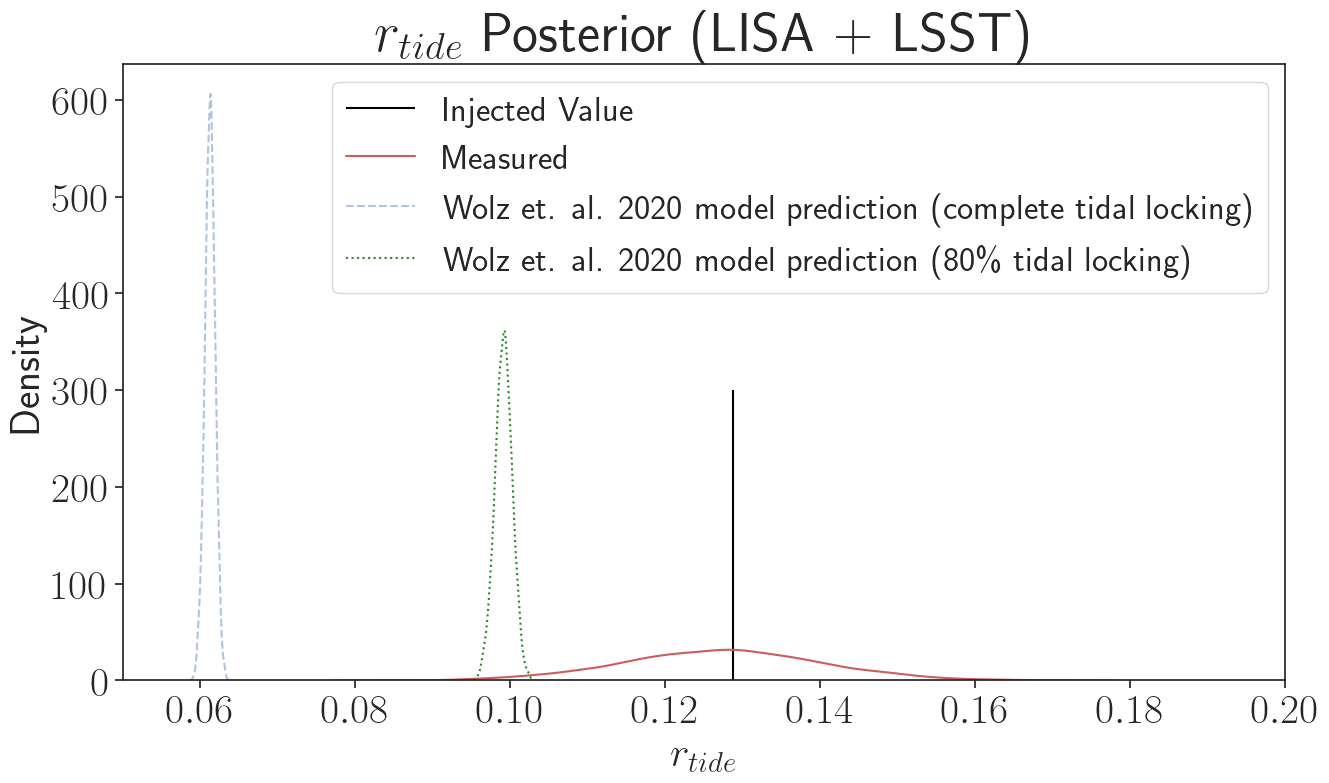}
    \caption{Multimessenger posteriors on $r_{\rm tide}$ with and without HiPERCAM data. In both plots, the rightmost, red, and solid curve shows our measured marginalized 1-D posterior for $r_{\rm tide}$. The leftmost, light blue, and dashed posterior uses the sampled masses to compute moment of inertia from the relation in \cite{Wolz_2020}, which is used to compute $r_{\rm tide}$ using Equations \ref{eq:r} and \ref{eq:rtide} with $\eta=1$. The center, green, and dotted posterior also uses the sampled masses to compute moment of inertia from \cite{Wolz_2020} to compute $r_{\rm tide}$ using Equations \ref{eq:r} and \ref{eq:rtide}, but with $\eta=0.8$. The total inconsistency with $\eta=1$ and evidence against $\eta=0.8$ demonstrates that the multimessenger approach can be sufficient to observationally constrain constrain the nonzero temperature effects on moment of inertia, even when we relax the assumption of complete tidal locking and only use LISA and LSST data.}
    \label{fig:zerotempIcomparison}
\end{figure}

The universal relations in \cite{Wolz_2020} predict a value of $r_{\rm tide}$ that is too low to be consistent with our injections, as can be seen by comparing the dashed blue and solid red posteriors in Figure \ref{fig:zerotempIcomparison}. Under our assumptions, including that the system is tidally locked ($\eta=1$), we could certainly rule this out and constrain temperature effects on moment of inertia. But if we were to relax the assumption of tidal locking ($\eta=1$), we would allow for larger values of $r_{\rm tide}$. From Figure 12 of \cite{fullerlai2012}, we can extrapolate the values of $\eta$ at a 6-min period to our WD temperatures of roughly 20,000 $\si{K}$ to estimate $\eta \approx 0.8$. We take this as a conservative assumption, as synchronization may be due to standing waves, which would suggest $\eta$ is closer to 1 than the traveling wave prediction.
We directly compare $r_{\rm tide}$ predictions from the zero-temperature star model in \cite{Wolz_2020} using both $\eta=1$ and $\eta=0.8$ with our inferred $r_{\rm tide}$ in Figure \ref{fig:zerotempIcomparison}. In the posterior for this example DWD system, we see complete inconsistency with the fully tidally locked predictions, and reasonably strong evidence against $\eta=0.8$. The evidence against $\eta=0.8$ is generally stronger when HiPERCAM data is included, but for this example DWD system, this evidence is dependent on the realization of the noise. From this, we conclude that multimessenger precision on $r_{\rm tide}$ can be sufficient to detect and constrain, in a way independent of WD stellar structure modeling, the degree of tidal locking and/or the nonzero temperature effects on moment of inertia.


Our posteriors were generated by sampling the physical parameters of the DWD system, which do not directly demonstrate our constraints on the phenomenological parameter $\ddot{f}$. Because $\ddot{f}$ is best constrained via EM measurement with high phase precision and many visits, we can estimate $\ddot{f}$ constraints using only HiPERCAM. We estimate the constraints on $\ddot{f}$ for this 6 min ZTF J2243+5242 source for varying lengths of observation times, frequencies of such observation times and total baselines. For each length of observation time, we run \texttt{pocoMC} on injected HiPERCAM data alone for a single segment of 3$\si{s}$ measurements for the given time and extract the marginalized uncertainty in the orbital phase. We use this same uncertainty for each observation time and fit a series of these with various frequencies and total baselines of these observations to a cubic polynomial and extract the uncertainty in $\ddot{f}$. The results are shown in Table \ref{tab:fddotconstraints}.

\addtolength{\tabcolsep}{20pt}
\begin{table*}[!]
\caption{\label{tab:fddotconstraints}
Uncertainties in the orbital frequency derivative $\ddot{f}$ computed by fitting orbital phase measurements of given visit frequencies to a cubic polynomial. For comparison, the standard deviation in the predicted orbital $\ddot{f}$ from the LISA + LSST + HiPERCAM posterior samples was $2.8\times 10^{-31}\,\,\si{s}^{-3}$. The left side value indicates the length of each observation chunk with a 3-second tme resolution and the top value indicates the frequency of such chunks. For example, the top left entry means that 1 hour of observation at the 3-second time resolution every year for 10 years (which corresponds to the cadence in Figure \ref{fig:cadences}) would yield an $\ddot{f}$ uncertainty of $4.8\times 10^{-28} \,\,\si{s}^{-3}$.}
\begin{ruledtabular}
\begin{tabular}{c|cccc}
& 10 years, yearly & 10 years, 10 times per year & 20 years, yearly & 20 years, 10 times per year  \\
\hline \\
1 hour & $1.1\times 10^{-27} \,\,\si{s}^{-3}$ & $4.3\times 10^{-28}\,\,\si{s}^{-3}$ & $1.1\times 10^{-28}\,\,\si{s}^{-3}$ & $3.9\times 10^{-29}\,\,\si{s}^{-3}$ \\
3 hours & $6.3\times 10^{-28}\,\,\si{s}^{-3}$ & $2.5\times 10^{-28}\,\,\si{s}^{-3}$ & $6.2\times 10^{-29}\,\,\si{s}^{-3}$ & $2.2\times 10^{-29}\,\,\si{s}^{-3}$ \\
5 hours & $4.6\times 10^{-28}\,\,\si{s}^{-3}$ & $1.8\times 10^{-29}\,\,\si{s}^{-3}$ & $4.5\times 10^{-29}\,\,\si{s}^{-3}$ & $1.6\times 10^{-29}\,\,\si{s}^{-3}$
\end{tabular}
\end{ruledtabular}
\end{table*}



\section{Discussion}
\label{sec:discussion}

Our analysis of this source demonstrates that the multi-messenger approach breaks the $\mathcal{M}$--$r_{\rm tide}$ degeneracy, while breaking this degeneracy using a measurement of $\ddot{f}$ is likely infeasible. For a number of reasons, we believe this analysis extends to other detached, tidally locked DWDs. As mentioned earlier, $\ddot{f} \propto f^{19/3}$, and we have selected a test case of very high orbital frequency. Our version of ZTF J2243+5242 has a shorter period than any other detached DWD system found so far \cite{7minDWD, gaialisa2018, gaialisa2024}, and this particular system is barely detached, so we believe this to be on the high end for $f$ and $\ddot{f}$. We expect lower frequency sources to have lower SNR in LISA, but the scaling is only $\rho_\text{LISA} \propto f^{10/3}$, so at frequencies where LISA SNR drops off to make multi-messenger degeneracy breaking infeasible, $\ddot{f}$ will have shrunk substantially more, making the EM approach even less feasible.

Some DWDs such as PG 1159-036 \cite{PG1159036} have been detected with at least one component exhibiting a surface temperature over 100,000 $\si{K}$, nearly an order of magnitude higher than that of ZTF J2243+5242, so one might consider the possibility of brighter sources yielding a much better phase measurement to measure $\ddot{f}$. The frequency constraint scales inversely with the SNR, which scales as $\rho \sim T^4\,f_b(T)$. For these hotter sources, these visible photometric bands are on the tail of the blackbody distribution, so we have $f_b(T) \sim T^{-3}$. This means $\delta\ddot{f} \sim T^{-1}$, which is not a scaling strong enough to make a substantial difference.

As shown in Figure 4, our $r_{\rm tide}$ constraints can be sufficient to distinguish from moment of inertia relations that neglect nonzero temperature effects. We should mention a few important caveats with this result. While the constraints seem to be stronger than the effects of our estimated uncertainty in $\eta$, the breakdown of the $\eta=1$ assumption would also open the door to a stronger impact of tidal heating, which cannot be completely neglected in \textit{nearly} tidally locked systems. Additionally, both the moment of inertia calculations of $\kappa_1=\kappa_2=0.12$ based on \cite{Burdge_2020} used in our injected value of $r_\textrm{tide}$ and the ``universal relations'' moment of inertia model in \cite{Wolz_2020} were for tidally unperturbed white dwarfs, which have slightly lower moments of inertia than those stretched by a companion. To first order, correcting for this would similarly shift both the mock measured and modeled posteriors for $r_{\textrm{tide}}$, so we would not expect the result to change, although it may be worthwhile to check this.

Temperature-dependent effects to these quasi-universal relations have been investigated in \cite{invalidILoveQ} and \cite{realisticILoveQ}. Multi-messenger observation could allow empirical testing of the deviations from these relations. Alternatively, if these relations were refined to accurately predict moment of inertia, we could instead use them to determine whether a tight DWD system is tidally locked.

\section{Conclusion}
\label{sec:conclusion}

We used MCMC to compute posteriors for a simulated Galactic eclipsing DWD system with mock LISA, LSST and HiPERCAM data that is similar to ZTF J2243+5242 with an orbital period of six minutes. The addition of LISA data allowed the breaking of degeneracies in mass, distance, radius, and $r_{\rm tide}$, the tidal contribution to the orbital frequency derivative $\dot{f}$. We demonstrate that masses and $r_{\rm tide}$ can be simultaneously measured to a precision that will allow us to constrain the non-zero temperature effects on the moments of inertia of the WDs. Without GW information from LISA, degeneracy breaking at a comparable level would not be possible even with dedicated 5-hour long HiPERCAM visits 10 times a year for 20 years. 

This degeneracy could be alternatively broken from dedicated spectroscopy, which could constrain radial velocity, but this would require valuable time on a very large telescope like the Extremely Large Telescope (ELT), the Giant Megellan Telescope (GMT), or the Thirty Meter Telescope (TMT) \cite{ELT}, which are likely to be severely oversubscribed. Hence, it is unclear whether spectroscopy could be efficient for studying a large sample of DWDs.
On the other hand, the multimessenger method we study in this work might identify many DWDs exhibiting interesting tidal effects on the orbital evolution, which will be justified follow-up targets for the very large telescopes. 

Multi-messenger analysis appears key to disentangling the impact of tides from the chirp masses. The effect of tides is the dominant deviation from the binary evolution prediction from solely GW radiation, so it will be important to measure this to accurately model the evolution of these DWDs as they approach merger. Constraining the tidal effects themselves will also allow us to empirically test stellar model predictions of moment of inertia. As these predictions improve, we would be able to constrain the degree of tidal locking $\eta$.
Joint efforts of the $\si{mHz}$-range GW community and the optical astronomy communities will be crucial for maximizing the science of Galactic DWDs in the next one or two decades.

\acknowledgments


NL thanks Quentin Baghi for assistance with developing his own LISA model code, Jakob Robnik for assistance with the planetary transit code for the eclipsing piece of the light-curve model, Minas Karamanis for help with sampling with \texttt{pocoMC}, Peter Yoachim for help with \texttt{rubin\_sim}, and Jim Fuller for helpful correspondence about DWD tides. L.D. acknowledges research grant support from the Alfred P. Sloan Foundation (Award Number FG-2021-16495), and support of Frank and Karen Dabby STEM Fund in the Society of Hellman Fellows.

\appendix
\section{Calibrations with HiPERCAM Observation on ZTF J2243+5242}
\label{sec:hipercamcalibrations}
We use our real HiPERCAM observation data of ZTF J2243+5242 to make our mock data as realistic as possible using our model. We perform a fit of the true data using our model to choose mock data parameters for our analysis and we perform a fit of the true noise estimates to calibrate our HiPERCAM noise.

\subsection{Choosing Mock Data Parameters using HiPERCAM Data}
\label{sec:hipercamdataparams}

The true HiPERCAM data is scaled relative to the brightness of a comparison star, so we do not use the data to model the temperature or luminosity distance, which scale the entire flux in each band. The data only lasts roughly an hour and a half, so we cannot reasonably use it to infer anything about the masses or $r_{\rm tide}$. So we fix the masses, temperatures, and luminosity distance according to the SED fit in \cite{8minDWD2020}, and fit phase, period, radii, inclination, irradiation amplitude, and flux scaling factors for each band.

\begin{figure}[!]
\centering
    \includegraphics[width=85mm]{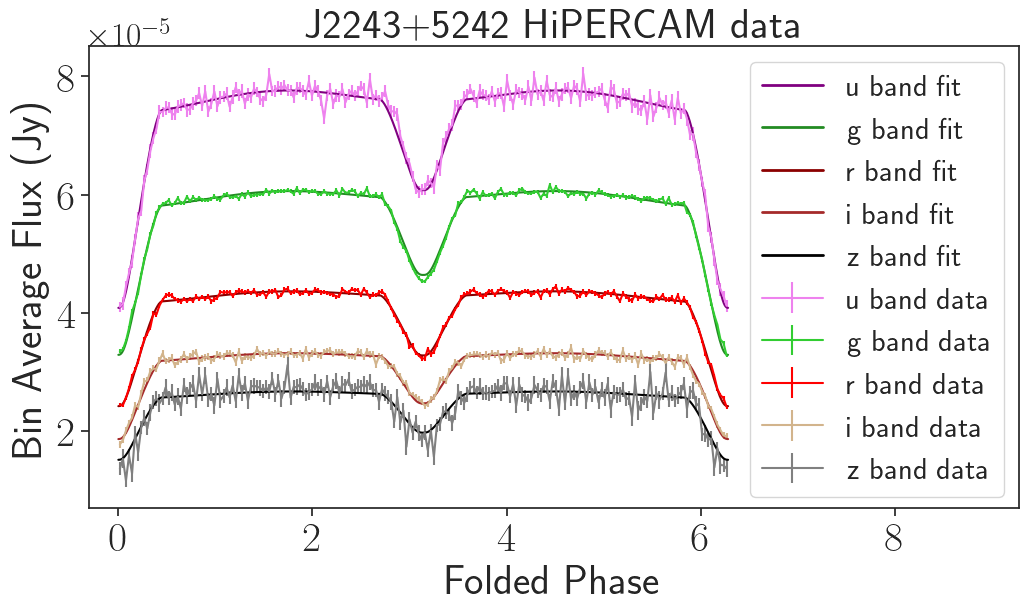} 
    \caption{Our best fit to a 1.5 hour HiPERCAM observation of ZTF J2243+5242. The data is phase folded using the best fit orbital phase parameters and averaged into 200 equally size phase bins. The data with error bars is shown as the lighter colors, and the model fit is the solid darker curve. The radii, inclination, and irradiation amplitude from this fit are used for our main analysis and are shown in Table \ref{tab:params}. The flux scaling factors are used to calibrate the mock noise, as detailed in section \ref{sec:hipercamnoise}.}
    \label{fig:hipercamdatafit}
\end{figure}

We compute posteriors in these parameters using the \texttt{MultiNest} sampler \cite{multinest}. Our best fit is shown in Figure \ref{fig:hipercamdatafit}. The radii, inclination, and irradiation amplitude in this fit are listed in Table \ref{tab:params}. Of note, the inclination is about 4.5 sigma above the result in \cite{8minDWD2020} and the radius of the secondary white dwarf is about 2.5 sigma below the result in \cite{8minDWD2020}. All other parameters are consistent. This difference is consistent with mild modeling differences that come from our simplifying assumptions, but we do not believe this will significantly affect the interpretation of our results.

\subsection{Calibrating Mock HiPERCAM Noise using HiPERCAM Data}
\label{sec:hipercamnoise}

\begin{figure}[!]
\centering
    \includegraphics[width=85mm]{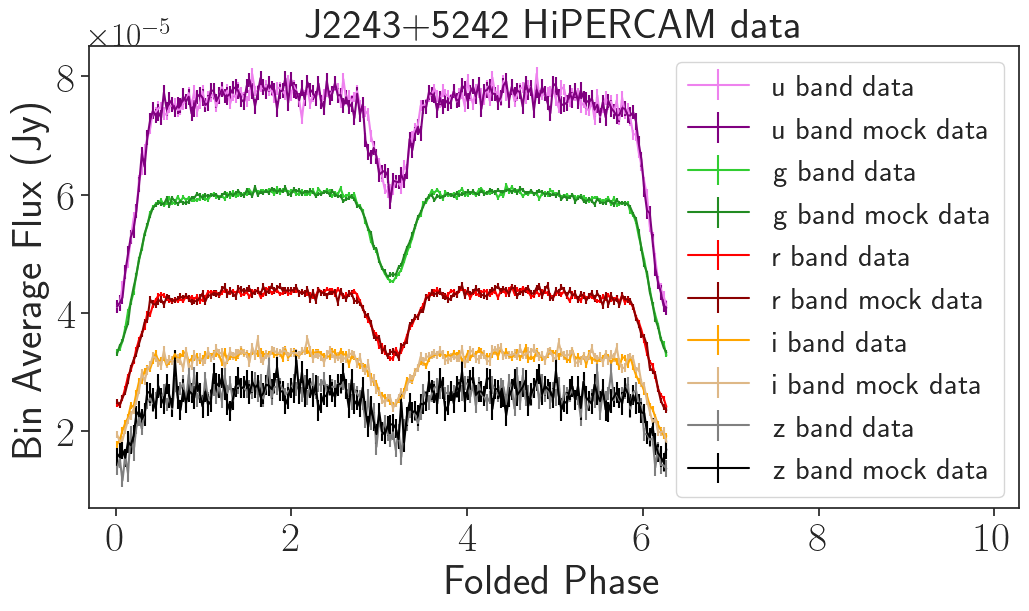} 
    \caption{A comparison between data from a real 1.5 hour HiPERCAM observation of ZTF J2243+5242 and mock data model with error bars from Equation \ref{eq:hipercamnoisemodel}. Both are phase folded using the best fit orbital phase parameters and averaged into 200 equally size phase bins. The true data is shown in lighter colors and the mock data is shown in darker colors.}
    \label{fig:hipercammockcompare}
\end{figure}

We used real HiPERCAM observations of ZTF J2243+5242 to calibrate our mock noise. For simplicity, we model our flux variance in each band $\sigma_b^2$ as a Poisson component plus a time-independent constant component: 

\begin{equation}
    \sigma_b^2 = \sigma_{\text{constant}, b}^2 + k_b \phi_b,
\label{eq:hipercamnoisemodel}
\end{equation}

where $\phi_b$ is the flux in band $b$, and $\sigma_{\text{constant}, b}$ and $k_b$ are the fit parameters. Noise from the real data had non-Poissonian time dependence due to factors like airmass, so the fit effectively averaged over these effects. A comparison between mock data using our best fit above and the fit noise and the true data is shown in Figure \ref{fig:hipercammockcompare}.

\section{Reparameterization for DWD Parameter Inference with Tides}
\label{sec:reparameterization}

Our model is parameterized by 14 total parameters: 13 physical parameters that we described in Table \ref{tab:params} (the individual masses $M_1$ and $M_2$ are replaced with the chirp mass $\mathcal{M}$ and mass ratio $q$) and an additional dummy mass ratio $q$. This parameterization has numerous degeneracies as seen in \ref{fig:posteriors}, which made it difficult or impossible for many samplers to converge. 

To solve this, we designed a reparameterization to remove the primary degeneracies, and this parameterization enabled us to get the accurate posteriors in \ref{fig:posteriors}. The $\mathcal{M}$/$r_{\rm tide}$ degeneracy suggested that we define a so-called "effective tidal chirp mass" or $\mathcal{M}_{\rm tide}$ as

\begin{equation}
    \mathcal{M}_{\rm tide} = \mathcal{M}(1 + r_{\rm tide})^{3/5}
\label{eq:Mtide}
\end{equation}

so that when you combine Equations \ref{eq:gwfdotpuregw} and \ref{eq:gwfdotrtide} you get $\dot{f} \propto \mathcal{M}_{\rm tide}^{5/3}$. The temperature, radii, and luminosity distance all contribute to the total fluxes from each star in the lightcurve model, so we address the resulting degeneracy by defining the coefficients $C_1$ and $C_2$ as

\begin{subequations}
\begin{equation}
    C_1 = \frac{T_1^4 R_1^2}{d_L^2},
\label{eq:C1}
\end{equation}
\begin{equation}
    C_2 = \frac{T_2^4 R_2^2}{d_L^2}.
\label{eq:C2}
\end{equation}
\label{eq:C1C2}
\end{subequations}

Finally we use the sum of the radii to make our parameter transformation bijective.

\begin{equation}
    R = R_1 + R_2
\label{eq:R}
\end{equation}

Our reparameterization is as follows:

\begin{equation}
    (r_{\rm tide}, d_L, R_1, R_2) \rightarrow (\mathcal{M}_{\rm tide}, R, C_1, C_2).
    \label{eq:reparam}
\end{equation}

To preserve the prior when sampling in this new space, one needs to add the log determinant of the Jacobian of the transformation to the log prior. If we call the original "physical" parameters  $\Theta_{\rm phys}$ and the new parameters that we use for sampling $\Theta_{\rm samp} = \mathcal{F}(\Theta_{\rm phys})$, then the determinant of that Jacobian is

\begin{equation}
    \det(J(\mathcal{F}^{-1}(\Theta_{\rm samp})) = \frac{5 d_L \mathcal{M}_{\rm tide}^{2/3}}{12\mathcal{M}^{5/3}C_1 C_2}\left( \frac{B_1}{D_2} + \frac{B_2}{D_1} \right),
\label{eq:detjac}
\end{equation}

where the coefficients $B$'s and $D$'s are defined as follows:

\begin{subequations}
\begin{equation}
    B_1 = \frac{K_1R}{D_1^2},
\label{eq:B1}
\end{equation}
\,
\begin{equation}
    B_2 = \frac{K_2R}{D_2^2},
\label{eq:B2}
\end{equation}
\begin{equation}
    D_1 = 1 + K_1,
\label{eq:D1}
\end{equation}
\begin{equation}
    D_2 = 1 + K_2,
\label{eq:D2}
\end{equation}
\begin{equation}
    K_1 = \sqrt{\frac{C_2}{C_1}}\left( \frac{T_1}{T_2}\right)^2,
\label{eq:K1}
\end{equation}
\begin{equation}
    K_2 = \sqrt{\frac{C_21}{C_2}}\left( \frac{T_2}{T_1}\right)^2.
\label{eq:K2}
\end{equation}
\label{eq:detjacdefs}
\end{subequations}

\bibliography{main}

\end{document}